\documentclass[aps,prl,twocolumn,superscriptaddress,english]{revtex4}
\usepackage{amssymb}
\usepackage{graphicx}
\usepackage{amsmath}
\usepackage{soul}
\usepackage{amsthm}
\usepackage{amsfonts}
\usepackage[T1]{fontenc}
\usepackage[latin9]{inputenc}
\usepackage{array}
\usepackage{multirow}
\usepackage{color}
\usepackage{esint}
\usepackage{bm}
\usepackage{color}
\usepackage{bbm}
\usepackage{hyperref}
\usepackage{babel}
\usepackage{titlesec}
\usepackage{braket}

\begin{document}

\global\long\def\id{\mathbbm{1}}
\global\long\def\ui{\mathbbm{i}}
\global\long\def\ud{\mathrm{d}}

%\title{Extending the concept of mobility edges; generalized}
\title{Experimental realization of the complete seven-phase Anderson-localization landscape}
 
\author{Yao Qin}
\altaffiliation{These authors contributed equally to this work.}
\affiliation{Department of Physics and State Key Laboratory of Quantum Functional Materials, Southern University of Science and Technology, Shenzhen 518055, China}
\affiliation{Shenzhen Institute for Quantum Science and Engineering and Guangdong Provincial Key Laboratory of Quantum Science and Engineering, Southern University of Science and Technology, Shenzhen 518055, China}

\author{Chao Yang}
\altaffiliation{These authors contributed equally to this work.}
\affiliation{Department of Physics and State Key Laboratory of Quantum Functional Materials, Southern University of Science and Technology, Shenzhen 518055, China}
\affiliation{Shenzhen Institute for Quantum Science and Engineering and Guangdong Provincial Key Laboratory of Quantum Science and Engineering, Southern University of Science and Technology, Shenzhen 518055, China}

\author{Yuzhe Zhang}
\affiliation{Department of Physics and State Key Laboratory of Quantum Functional Materials, Southern University of Science and Technology, Shenzhen 518055, China}
\affiliation{Shenzhen Institute for Quantum Science and Engineering and Guangdong Provincial Key Laboratory of Quantum Science and Engineering, Southern University of Science and Technology, Shenzhen 518055, China}

\author{Yucheng Wang}
\email{wangyc3@sustech.edu.cn}
\affiliation{Shenzhen Institute for Quantum Science and Engineering and Guangdong Provincial Key Laboratory of Quantum Science and Engineering, Southern University of Science and Technology, Shenzhen 518055, China}

\author{Jingyun Fan}
\email{fanjy@sustech.edu.cn}
\affiliation{Department of Physics and State Key Laboratory of Quantum Functional Materials, Southern University of Science and Technology, Shenzhen 518055, China}
\affiliation{Shenzhen Institute for Quantum Science and Engineering and Guangdong Provincial Key Laboratory of Quantum Science and Engineering, Southern University of Science and Technology, Shenzhen 518055, China}
\affiliation{Center for Advanced Light Source, Southern University of Science and Technology, Shenzhen, 518055, China}
\affiliation{Hefei National Laboratory, Hefei 230088, China}
%\date{\today}

\begin{abstract}

Anderson localization has evolved far beyond the conventional dichotomy between extended and localized states. Modern localization theory predicts a complete transport hierarchy comprising extended, critical, and localized phases together with all coexistence phases among them, forming a seven-phase Anderson-localization landscape. Despite its fundamental importance, this hierarchy has never been experimentally realized within a single system. Here we realize the complete seven-phase Anderson-localization landscape in a one-dimensional Floquet photonic lattice. By engineering quasiperiodic hopping profiles containing inhomogeneously distributed hopping zeros, we generate critical states and enable their coexistence with extended and localized sectors. The resulting transport regimes are directly resolved through their distinct spatiotemporal dynamics, including ballistic expansion, confined critical oscillations, and persistent localization. We observe all seven phases, including the elusive triply coexisting extended-critical-localized phase, and experimentally track the phase transitions connecting them. Our results establish the first complete experimental map of the Anderson-localization landscape and provide a unified platform for investigating mobility edges, multifractality, and programmable coherent transport.

\end{abstract}
\maketitle

\noindent \textbf{Introduction.}
Since the seminal discovery by Philip W. Anderson in 1958 that disorder can completely suppress quantum transport~\cite{Anderson1}, Anderson localization has become a foundational paradigm of wave physics~\cite{Anderson2,Anderson3,Anderson4,Andersonexp1,Andersonexp2,Andersonexp3,Andersonexp4,Andersonexp5,Andersonexp6,Andersonexp7,Andersonexp8,Andersonexp9}. Over the past decades, its understanding has evolved far beyond the original extended--localized dichotomy. What was once regarded as a singular critical point---the mobility edge (ME) separating extended ($E$) and localized ($L$) states~\cite{Anderson2,Anderson3}---is now understood to support finite regions of critical ($C$) states characterized by multifractal
wave functions, singular continuous spectra, 
and anomalous transport intermediate between ballistic expansion and exponential localization~\cite{critical2,critical3,critical4,critical5,critical6,critical7,critical8,critical9,criticalexp1,criticalexp2,criticalexp3,criticalexp4,criticalexp5}. Together with their coexistence phases, these three transport sectors define a complete Anderson-localization landscape comprising seven fundamentally distinct phases.

The emergence of this richer transport hierarchy has transformed the understanding of wave dynamics across condensed-matter physics, photonics, acoustics, superconducting circuits, and ultracold atomic platforms~\cite{Anderson4,Andersonexp1,Andersonexp2,Andersonexp3,Andersonexp4,Andersonexp5,Andersonexp6,Andersonexp7,Andersonexp8,Andersonexp9,critical9,criticalexp1,criticalexp2,criticalexp3,criticalexp4,criticalexp5,AAexp,AAexp2,ME6,ME7,ME8,ME9,ME10,ME12}, while providing a versatile framework for exploring ergodicity breaking and nonequilibrium dynamics~\cite{MBL1,MBL2}, as well as for engineering programmable wave propagation~\cite{Anderson4,Andersonexp1,Andersonexp2,Andersonexp3,Andersonexp4,Andersonexp5,Andersonexp6,Andersonexp7,Andersonexp8,Andersonexp9,critical9,criticalexp1,criticalexp2,criticalexp3,criticalexp4,criticalexp5,AAexp,AAexp2,ME6,ME7,ME8,ME9,ME10,ME12}. A complete experimental realization of this hierarchy within a single controllable system would provide a long-sought opportunity to explore the full Anderson-localization landscape, the interplay between distinct transport sectors, the ME phenomena and transport-phase transitions. Yet despite decades of theoretical and experimental advances, the simultaneous coexistence of the three fundamental classes has remained experimentally elusive, let alone the realization of the complete seven-phase hierarchy within a single quantum system. 

Quasiperiodic systems have played a particularly important role in this conceptual evolution~\cite{critical2,critical3,critical4,critical5,critical6,critical7,critical8,critical9,criticalexp1,criticalexp2,criticalexp3,criticalexp4,criticalexp5,AAexp,AAexp2,ME6,ME7,ME8,ME9,ME10,ME12,ME1,ME2,ME3,ME4,ME5}. In contrast to conventional disordered systems---where infinitesimal uncorrelated disorder localizes all states in one and two dimensions~\cite{Abrahams1979}, and MEs separating extended and localized states arise only in sufficiently disordered three-dimensional systems---quasiperiodic lattices can host the complete hierarchy of localization phases already within one dimension~\cite{critical2,critical3,critical4,critical5,critical6,critical7,critical8,critical9,criticalexp1,criticalexp2,criticalexp3,criticalexp4,criticalexp5,AAexp,AAexp2,ME6,ME7,ME8,ME9,ME10,ME12,ME1,ME2,ME3,ME4,ME5}. More recently, the application of Avila's Global Theory to quasiperiodic operators has revealed an elegant mechanism that critical states emerge as dynamically caged modes confined between inhomogeneously distributed hopping zeros (IDZs)~\cite{critical7,critical8,criticalexp4,criticalexp5,Avila2015}. Here, by engineering a Floquet photonic lattice incorporating such IDZ structures, we realize the first complete experimental map of the seven-phase Anderson localization landscape---including the elusive triply coexisting ($E+L+C$) phase---within a single nearest-neighbor lattice.

\noindent \textbf{Floquet Model.} 
We study a one-dimensional spinful Floquet lattice of length $N$ with nearest-neighbor hopping. The single-period evolution under open boundary conditions is governed by a sequence of four non-commuting unitary operations,
\begin{equation}
    \mathcal{F}_{\rm OBC} = U_4 U_3 U_2 U_1.
\end{equation}
The first step, $U_1=\sum_{n=1}^{N}\ket{n}\bra{n}\otimes O(\theta_n)$, applies site ($n$)-dependent spin rotations in the two-component internal spin space $(\uparrow,\downarrow)$, where $O(\theta_n)=e^{-{\mathrm{i}}\theta_n\sigma_y}$. Spatial transport is generated by two translation-assisted spin rotations, $U_2=S^{-1}[\ket{1}\bra{1}\otimes\mathbb{I}
+\sum_{n=2}^{N}\ket{n}\bra{n}\otimes O(\theta_{n-1}^{(+)})]S$ and $U_3=S[\sum_{n=1}^{N-1}\ket{n}\bra{n}\otimes O(-\theta_{n}^{(-)})
+\ket{N}\bra{N}\otimes\mathbb{I}]S^{-1}$. Because the spin-dependent translation operator \(S=\sum_{n=1}^{N}(\ket{n+1,\uparrow}\bra{n,\uparrow}
+\ket{n,\downarrow}\bra{n,\downarrow})\) shifts only the $\uparrow$-spin component, internal spin rotations are converted into directional nearest-neighbor hopping. Finally, $U_4=\sum_{n=1}^{N}\ket{n}\bra{n}\otimes
e^{i(-1)^{n+1} \varphi \sigma_z}$ introduces spin-dependent, staggered onsite phases. The effective Hamiltonian of this Floquet system is schematically shown in Fig. 1a (see Appendix).

To engineer the complete seven-phase landscape within a single nearest-neighbor lattice, we impose a quasiperiodic modulation on the rotation angles. Specifically, we choose $\theta_n=(-1)^n\frac{J}{2}$, while $\theta_{n}^{(\pm)}=J+\lambda_{\pm} J\cos[2\pi\beta(n+1)] $ on odd sites and $(5.2-1.3\lambda_{\mp})J$ on even sites with \(\beta=(\sqrt{5}-1)/2\). The staggered onsite phase strength is chosen as 
$\varphi=[10.4-1.3(\lambda_{+}+\lambda_{-})]J$. The two control parameters $(\lambda_{+},~\lambda_{-})$ tune the quasiperiodic hopping profile and the formation of IDZs. These zeros act as deterministic transport barriers that partition the lattice into dynamically isolated regions. According to Avila's Global Theory, the resulting caging mechanism stabilizes critical states and enables their coexistence with extended and localized sectors within a single spectrum~\cite{critical7,critical8,criticalexp4,criticalexp5,Avila2015} (see Supplemental Information). 

To characterize the resulting localization properties, we diagonalize the Floquet operator and compute the inverse participation ratio and fractal dimension ($D$) for each eigenstate (see Appendix). In the thermodynamic limit, $D=1,~0$, and $0<D<1$ correspond to  extended, localized, and critical states, respectively. Finite-size scaling of $D$ yields the global phase diagram and determines the boundaries between all seven phases. As shown in Fig.~1b, varying $(\lambda_+,~\lambda_-)$ at fixed \(J=0.2\) drives the system through the complete seven-phase localization landscape. 

\begin{figure}[t]
	\centering
	\includegraphics[width=1.0\columnwidth]{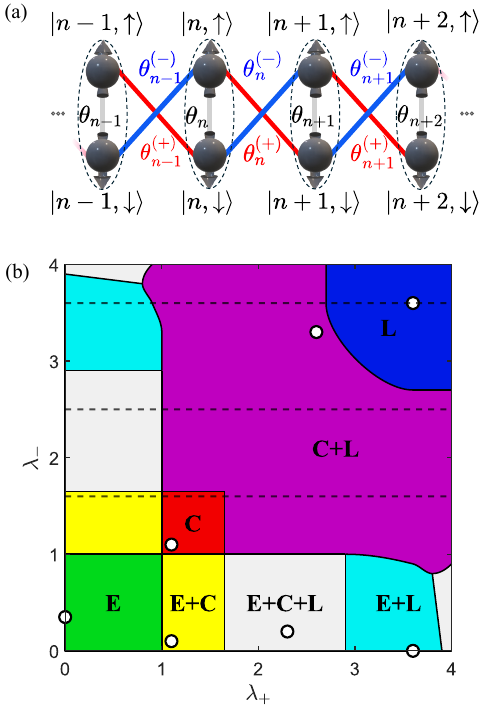}
\caption{
\textbf{Complete seven-phase Anderson-localization landscape.}
\textbf{a,} Schematic of the effective Floquet Hamiltonian,
which consists of vertical bonds (onsite spin-flip couplings $\theta_n$), diagonal bonds (translation-assisted couplings $\theta_n^{+}$ and $\theta_n^{-}$), and a staggered spin-dependent onsite potential. 
\textbf{b,} Phase diagram in the control-parameter space $(\lambda_{+},\lambda_{-})$ obtained from finite-size scaling of the fractal dimension $D$. It comprises three pure phases: extended ($E$), critical ($C$), and localized ($L$), and four coexistence phases: $E+L$, $C+L$, $E+C$, and $E+C+L$. Open circles denote the experimentally probed parameters in Fig.~3 and black dashed lines mark the phase-transition paths studied in Fig.~4. 
}
	\label{fig1}
\end{figure}

\begin{figure}[t]
	\centering
	\includegraphics[width=1.0\columnwidth]{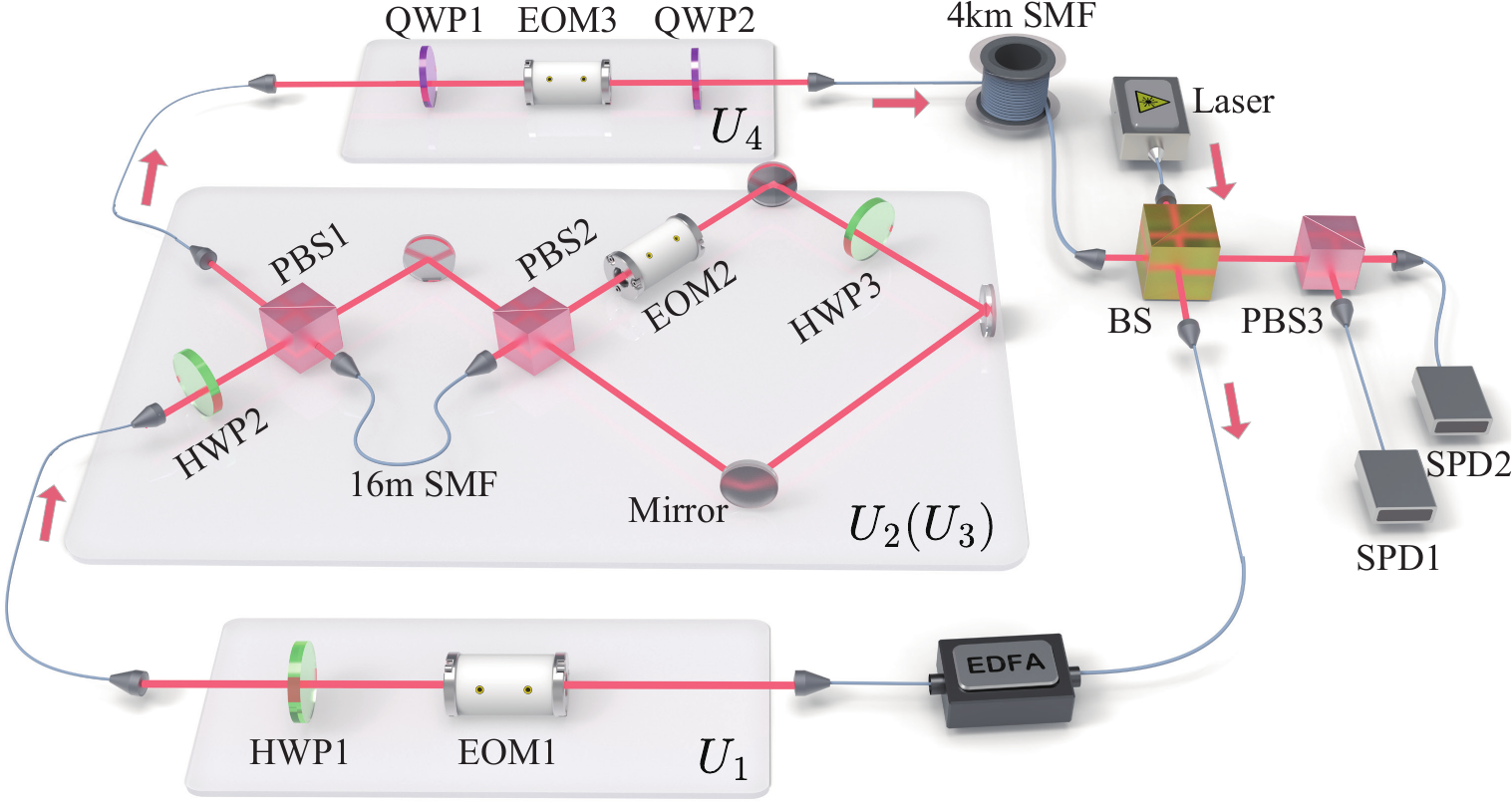}
\caption{
\textbf{Experimental implementation of the Floquet lattice in a recirculating optical-loop architecture.}
A 6-ns laser pulse (1560 nm) circulates in a fiber loop to emulate the Floquet operator $\mathcal{F}_{\rm OBC}=U_4U_3U_2U_1$. One Floquet cycle is implemented over two successive round trips (see Appendix): the first implements $U_1$ and $U_2$ and the second implements $U_3$ and $U_4$. 
The onsite spin rotation $O(\theta_n)$ is implemented by a half-wave plate (HWP1) and an electric optical modulator (EOM1). The hopping operations $U_2$ and $U_3$ are realized using a polarization-dependent double-loop interferometer, where a 16-m single-mode fiber introduces an 80-ns delay between the $|H\rangle$ and $|V\rangle$ polarization components, implementing the shift operator $S$, while the return path realizes $S^{-1}$. Unlike $U_2$, which applies $S$ followed by $S^{-1}$, the operation $U_3$ performs $S^{-1}$ followed by $S$; this distinction is achieved by inserting a HWP2 oriented at $45^\circ$. Site-dependent spin rotations $O(\theta_n^{(+)})$ and $O(-\theta_n^{(-)})$ are encoded by EOM2 and HWP3. The staggered onsite phase $U_4$ is generated by EOM3 placed between quarter-wave plates QWP1 and QWP2. A 4-km fiber prevents temporal overlap between successive round trips. After each round trip, $10\%$ of the photons are extracted by a beam splitter (BS), pass a polarizing beam splitter (PBS), and are detected by single-photon detectors (SPDs). An erbium-doped fiber amplifier (EDFA) compensates for losses. Measurement statistics are recorded after every second round trip, corresponding to one complete Floquet cycle. 
}
	\label{fig2}
\end{figure}

\begin{figure*}[t]
	\centering
	\includegraphics[width=1\textwidth]{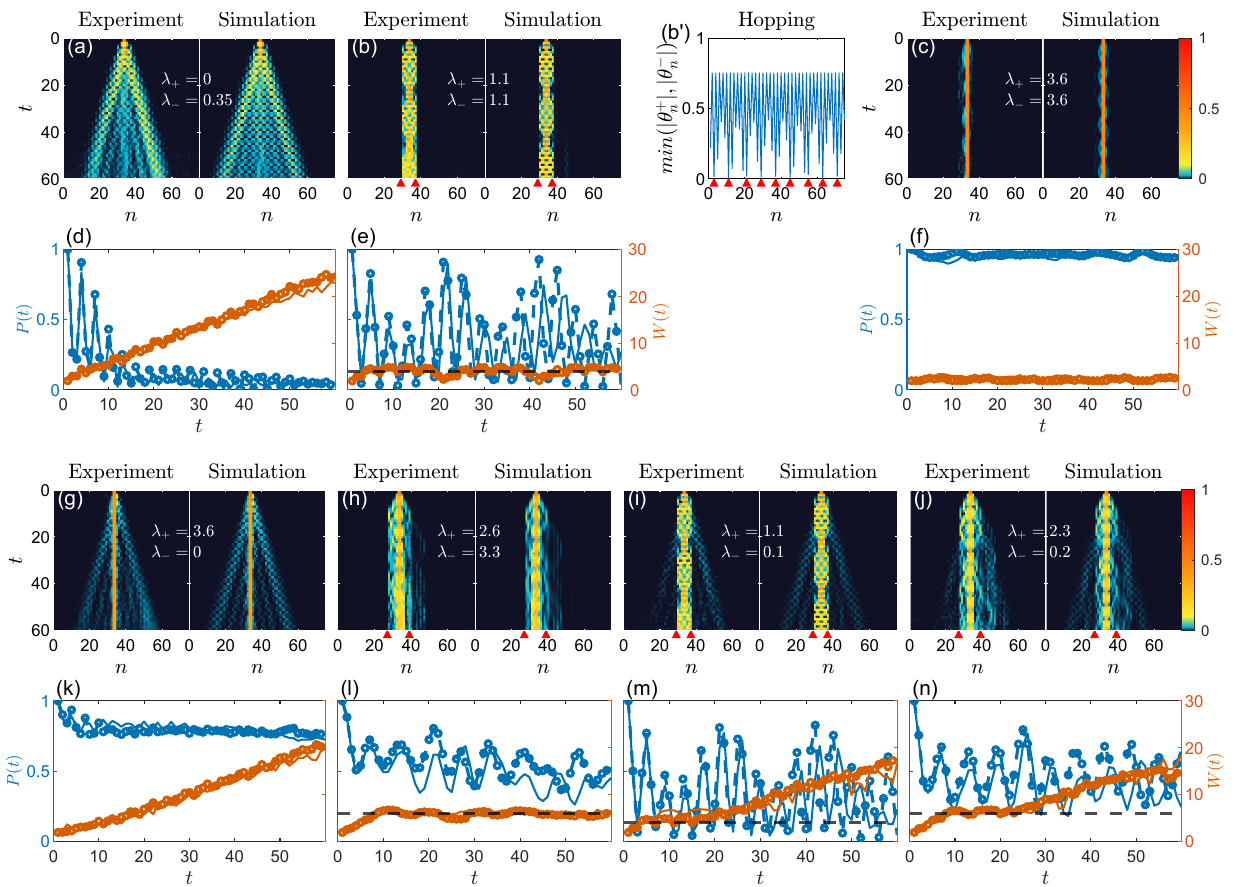}
	\caption{
\textbf{Experimental realization of the complete seven-phase Anderson-localization landscape}. The initial state is $|\psi(0)\rangle=(2|n_0,\uparrow\rangle+|n_0+1,\uparrow\rangle+2i|n_0+1,\downarrow\rangle)/3$ with $n_0=33$ and the evolution is monitored for $T=60$ Floquet cycles. 
\textbf{Upper panels:}  
\textbf{a--c}, Measured spatiotemporal distributions and corresponding numerical simulations for the three fundamental transport regimes: (\textbf{a}) extended ($E$), (\textbf{b}) critical ($C$), and (\textbf{c}) localized ($L$). The color scale represents $p(n,t)$.
\textbf{b$'$}, Hopping landscape for the critical phase in \textbf{b}, with red arrows indicating the positions of the inhomogeneously distributed hopping zeros (IDZs). 
\textbf{d--f}, Corresponding quantitative diagnostics: wavefront $W(t)$ and survival probability $P(t)$. 
Open symbols connected by dashed lines denote experimental measurements, solid lines denote numerical simulations, and black dashed line in \textbf{e} indicate the IDZ-defined confinement boundaries.  
\textbf{Bottom panels:} 
\textbf{g--j}, Measured spatiotemporal distributions and corresponding numerical simulations of the four coexistence phases: (\textbf{g}) $E+L$, (\textbf{h}) $C+L$, (\textbf{i}) $E+C$, and (\textbf{j}) $E+C+L$. 
\textbf{k--n}, Corresponding wavefront $W(t)$ and survival probability $P(t)$. 
Red arrows denote the positions of the IDZs, while black dashed lines indicate the corresponding IDZ-defined confinement boundaries separating the extended from the critical and localized components, where applicable. 
    }
	\label{fig3}
\end{figure*}

%%%%%%%%%%%%%%%%%%%%%%%%%%%%
\noindent \textbf{Experimental setup.}  
Fig.~\ref{fig2} schematically illustrates the experimental realization of the Floquet model using a recirculating optical-loop architecture. Coherent optical propagation follows the same unitary dynamics as a single quantum particle, enabling laser-pulse evolution to faithfully emulate the Floquet lattice~\cite{PQW1,PQW2,PQW3,PQW4,PQW5,PQW6}. In this implementation, horizontal and vertical polarizations encode the synthetic spin states $\uparrow$ and $\downarrow$, while lattice sites are mapped onto discrete time bins~\cite{PQW5,PQW6}.  Notably, the Floquet operator $\mathcal{F}_{\rm OBC}=U_4U_3U_2U_1$ is synthesized over two successive round trips. The first round trip implements $U_1$ and $U_2$, whereas the second implements $U_3$ and $U_4$ (see Appendix). Upon completion of each Floquet cycle, $10\%$ of the pulse energy is extracted for time-resolved detection. The measured spatiotemporal distribution 
$p(n,t)=\big\lvert\langle n | \mathcal{F}_{\rm OBC}^{\,t} \,|\psi(0)\rangle\big\rvert^2$ 
directly maps the site-resolved Floquet dynamics of the synthetic lattice, where $t$ denotes the number of completed Floquet cycles, and $p(n,t)=p(n,\uparrow,t)+p(n,\downarrow,t)$ accounts for photons in both vertical and horizontal polarizations.

\noindent \textbf{Three fundamental transport regimes.} 
To experimentally benchmark the three fundamental transport regimes, we initialize the pulse in the state $|\psi(0)\rangle=(2|n_0,\uparrow\rangle+|n_0+1,\uparrow\rangle+2i|n_0+1,\downarrow\rangle)/3$ with $n_0=33$ in a 75-site synthetic lattice and monitor its evolution for $T=60$ Floquet cycles. Figs.~3a-3c present measured spatiotemporal distributions alongside numerical simulations for three representative parameter sets: $(\lambda_+,\lambda_-)=(0,0.35)$ (extended), $(1.1,1.1)$ (critical), and $(3.6,3.6)$ (localized), selected from the phase diagram in Fig.~1. 
From left to right, the measured distributions exhibit a clean symmetric expansion cone (Fig.~3a), indicating ballistic transport across the lattice; bounded spreading within an IDZ-defined region accompanied by persistent oscillatory dynamics (Fig.~3b), characteristic of the critical phase; and strong confinement near the initial excitation site (Fig.~3c), indicative of localization~\cite{criticalexp4,criticalexp5}. These distinct spatiotemporal signatures provide direct experimental identification of the extended, critical, and localized transport regimes. Corresponding to Fig.~3b, Fig.~3b$'$ displays the  hopping profile, highlighting the IDZs that partition the lattice into dynamically isolated regions and thereby generate the caged oscillatory dynamics characteristic of the critical phase.

To quantify these transport behaviors, we introduce two complementary observables (see Supplementary Information): the survival probability $P(t)$ and the quantile wavefront $W(t)$. The survival probability is defined as the total excitation probability within the initially occupied region $\mathcal{C}_0$ (the two neighboring sites $n_0$ and $n_0+1$): $ P(t) = \sum_{n\in\mathcal{C}_0} p(n,t). $ This observable directly probes the level of localization. 
The quantile wavefront $W(t)$ is defined as the distance from the initial excitation region $\mathcal{C}_0$ to the position at which the cumulative probability in the propagating tail reaches $\eta$. Throughout this work, we fix $\eta=5\%$, thereby probing the most mobile component of the wavepacket.

As shown in Figs.~3d-3f, the three regimes exhibit sharply distinct dynamical responses. In the extended regime (Fig.~3d), $W(t)$ grows linearly with time, $W(t)\propto t$, while $P(t)$ rapidly decays toward zero. In the critical regime (Fig.~3e), $W(t)$ initially increases before saturating at a finite value ($W\approx4$), coinciding with the IDZ-defined confinement boundaries indicated by the black dashed line, whereas $P(t)$ remains finite and exhibits persistent irregular oscillations, reflecting ongoing probability redistribution within an IDZ-confined support. In the localized regime (Fig.~3f), $W(t)$ remains strongly suppressed $(W\lesssim2)$ throughout the evolution, while $P(t)$ stays close to unity, indicating sustained localization. The remarkable agreement between experiment (open circles) and theory (smooth lines) confirms the realization of all three fundamental transport regimes and validates the underlying IDZ-based mechanism responsible for the emergence of critical states.

\noindent \textbf{Four coexistence phases.} 
Having established the three fundamental transport regimes, we next explore the four coexistence phases that complete the seven-phase Anderson-localization landscape. These hybrid regimes arise when MEs partition the Floquet spectrum into sectors with distinct transport properties, allowing extended, critical, and localized eigenstates to coexist within a single parameter configuration.

The bottom panels of Fig.~3 extend the experimental study to the four coexistence phases, using the same initial state as in the three fundamental transport regimes. In the $E+L$ regime (Fig.~3g), a sharply localized central ridge coexists with a linearly expanding transport cone. This coexistence is reflected by a finite plateau in $P(t)$, originating from the localized sector, together with persistent linear growth of $W(t)$ driven by the extended states (Fig.~3k). 
In the $C+L$ regime (Fig.~3h), a localized peak is superimposed on an IDZ-confined oscillatory structure. Here $W(t)$ rapidly saturates at a finite value $(W\approx6)$, while $P(t)$ remains high and exhibits persistent irregular oscillations (Fig.~3l), demonstrating the coexistence of localization and critical dynamics in the absence of propagating transport channels. In the $E+C$ regime (Fig.~3i), a bounded oscillatory core associated with critical dynamics is embedded within a broad ballistic background. Correspondingly, $P(t)$ decays to a low but strongly fluctuating value, while $W(t)$ continues to grow linearly (Fig.~3m), indicating the simultaneous presence of critical confinement and ballistic transport.

The most intricate behavior appears in the $E+C+L$ regime (Fig.~3j), where all three transport sectors coexist simultaneously. The spatiotemporal distribution resolves a localized central peak, a bounded critical cage, and a weak ballistic background extending across the lattice. This three-component structure generates a unique dynamical fingerprint: $W(t)$ displays two-stage evolution, with an initial saturation associated with IDZ-constrained critical spreading followed by linear growth arising from the extended sector, while $P(t)$ combines a finite localized baseline with large-amplitude critical oscillations (Fig.~3n). These observations constitute the first experimental realization of the triply coexisting $E+C+L$ phase.

Collectively, the results of Fig.~3 provide the first complete experimental map of the seven-phase Anderson-localization landscape. The distinct dynamical fingerprints of each regime enable unambiguous discrimination of all transport phases and show excellent agreement with theoretical predictions.

\begin{figure*}[t]
	\centering
	\includegraphics[width=1\textwidth]{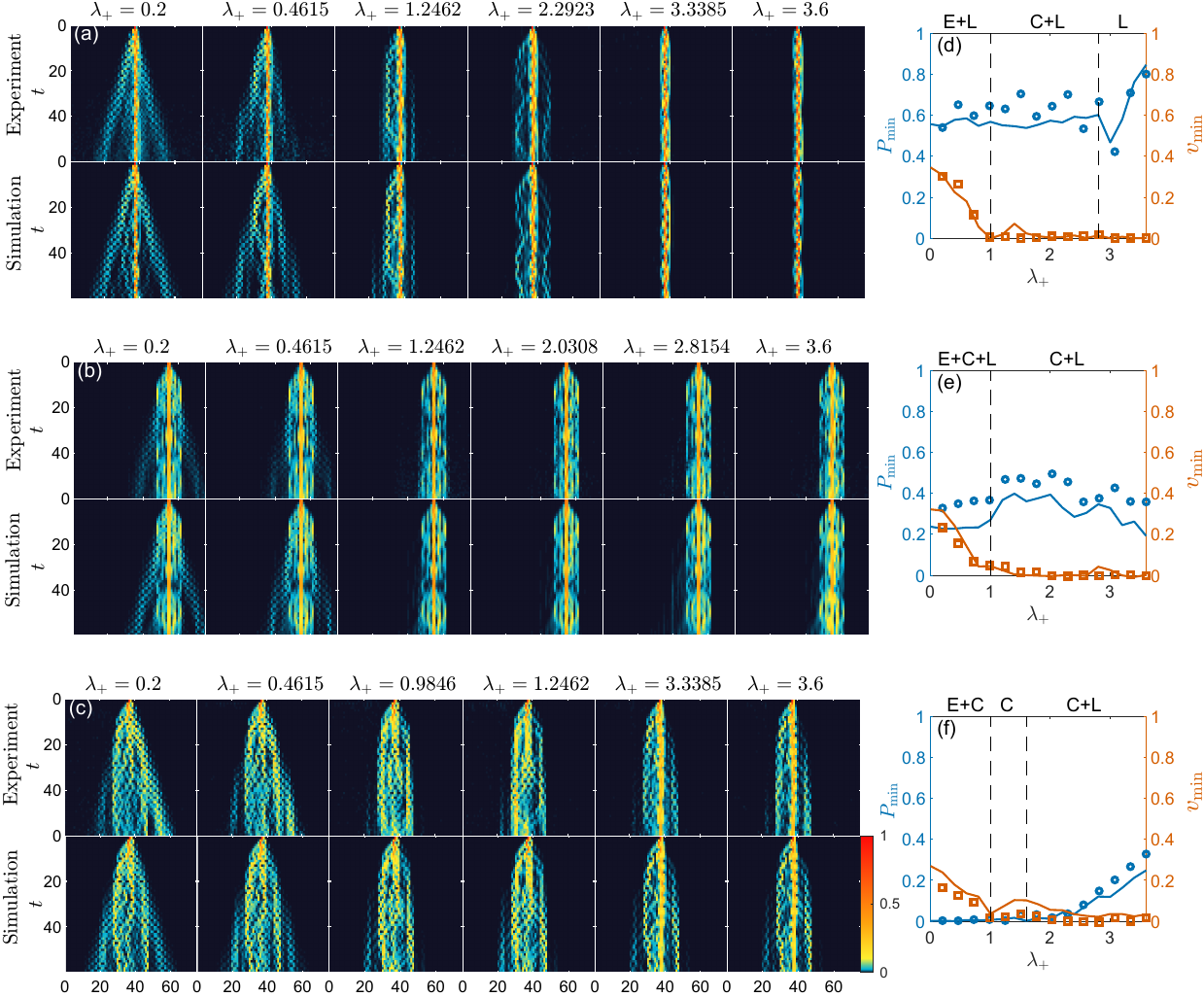}
\caption{
\textbf{Phase transitions across the seven-phase Anderson-localization landscape.} 
Left panels \textbf{a--c}, measured and simulated spatiotemporal dynamics obtained by varying $\lambda_+$ at fixed $\lambda_-$ along the trajectories indicated in Fig.~1. The color scale represents $p(n,t)$. \textbf{a}: transition $E+L \rightarrow C+L \rightarrow L$, with initial state $|\psi(0)\rangle=0.5234|n_0,\uparrow\rangle+0.8375|n_0,\downarrow\rangle-0.1570i|n_0+1,\uparrow\rangle$, $n_0=37$ and $\lambda_-=3.6$. 
\textbf{b}: transition $E+C+L \rightarrow C+L$, $|\psi(0)\rangle=0.6509|n_0,\uparrow\rangle+0.3906|n_0,\downarrow\rangle-0.6509i|n_0+1,\downarrow\rangle$, $n_0=54$ and $\lambda_-=2.5$. 
\textbf{c}: transition $E+C \rightarrow C \rightarrow C+L$, $|\psi(0)\rangle=0.4149|n_0,\uparrow\rangle+0.8297|n_0,\downarrow\rangle-0.3734i|n_0+1,\downarrow\rangle$, $n_0=37$ and $\lambda_-=1.6$. 
The evolving spatiotemporal distribution reveal the disappearance and emergence of extended, critical, and localized transport sectors across the phase boundaries. 
Right panels \textbf{d--f}, quantitative characterization of the corresponding transitions. Blue circles (experimental) and lines (numerical) denote the survival probability $P_{\min}$, while orange squares (experimental) and lines (numerical) denote the wavefront velocity $v_{\text{min}}$. Vertical dashed lines indicate the phase boundaries obtained from spectral analysis (see Supplementary Information).
}
	\label{fig4}
\end{figure*}

\noindent \textbf{Phase transitions.} 
Lastly, we investigate phase-transition trajectories corresponding to the black dashed lines in Fig.~1. The top, middle, and bottom panels of Fig.~4 trace the transitions $E+L \rightarrow C+L \rightarrow L$, $E+C+L \rightarrow C+L$, and $E+C \rightarrow C \rightarrow C+L$, respectively. Along these trajectories, the spatiotemporal dynamics reveals a systematic reorganization of transport sectors. In the first case (Fig.~4a), the dynamics evolves from ballistic-localized coexistence to critical-localized coexistence and ultimately to complete localization. In the second (Fig.~4b), a trimodal distribution containing extended, critical, and localized components progressively loses its ballistic sector, leaving only critical-localized coexistence. In the third (Fig.~4c), the ballistic background is gradually suppressed, driving the system from extended-critical coexistence to a purely critical phase before localized states emerge and coexist with the critical sector. In each case, the initial state is chosen to maximize the visibility of the evolving dynamical signatures as the control parameter $\lambda_+$ is varied. 

The transitions are quantitatively characterized by two observables: the wavefront velocity $v_{\min}$ and the minimum survival probability $P_{\min}$. Specifically, the wavefront velocity is extracted from the quantile wavefront $W(t)$ by linear fitting over a sliding time window $t\in[t_0,t_0+20]$, $W^{\text{fit}}(t) \simeq v t + b$, yielding $v = dW^{\text{fit}}/dt$. We then define $v_{\min}=\min_{t_0}|v|$, which quantifies the slowest propagation rate during the evolution. The minimum survival probability is defined as $P_{\min} = \min_{0\leq t\leq T} P(t)$. These quantities are shown in the right panels of Fig.~4. 
Across all trajectories, $v_{\min}$ decreases and vanishes as the extended sector disappears, providing a direct signature of the suppression of ballistic transport. In contrast, $P_{\min}$ remains finite whenever localized states are present, reflecting a nonzero localized fraction that prevents complete depletion of the initial-site occupation. As the system approaches the fully localized regime, $P_{\min}$ increases toward unity. The $C+L$ phase is therefore identified by a vanishing wavefront velocity together with finite $P_{\min}$, indicating the coexistence of critical and localized sectors in the absence of extended transport channels. By comparison, the pure critical phase is characterized by vanishing $v_{\min}$ together with nearly vanishing $P_{\min}$, indicating the absence of both extended and localized sectors while preserving IDZ-confined non-ergodic dynamics.

The calculated phase boundaries (vertical dashed lines), determined solely from the spectral fractions of extended, critical, and localized eigenstates, agree closely with the experimentally observed variations of $v$ and $P_{\min}$, providing direct validation of the predicted phase diagram. The only noticeable deviation occurs in Fig.~4f. This mismatch originates from the fact that the theoretical phase boundaries are extracted purely from the eigenstate composition of the spectrum, whereas the experiment probes only those states significantly populated by the chosen initial condition. In this case, the overlap of the initial state with the localized sector is weak, leading to reduced experimental sensitivity to the transition and a displacement of the phase boundary (see Supplementary Information). Although the combination of $v_{\min}$ and $P_{\min}$ provides a means of experimentally distinguishing all seven transport regimes, a complete quantitative characterization of criticality---particularly in finite systems where multifractal scaling is only partially developed---remains an open challenge. The development of experimental observables capable of directly probing critical dynamics and multifractal structure therefore represents an important direction for future research. The agreement between experiment and theory confirms both the predicted phase transitions and the realization of the complete seven-phase Anderson-localization landscape within a single controllable Floquet platform.

\noindent \textbf{Conclusion.} 
In summary, we have experimentally realized the complete seven-phase Anderson-localization landscape within a single nearest-neighbor Floquet lattice. By engineering quasiperiodic hopping profiles that generate inhomogeneously distributed zeros, we directly observed extended, critical, and localized transport sectors together with all four coexistence phases featuring mobility edges. The measured spatiotemporal dynamics and transport observables provide a unified framework for identifying all localization regimes and tracking the phase transitions connecting them. Our work establishes a versatile platform for engineering and exploring the full hierarchy of localization physics within a single controllable system. The ability to create, manipulate, and interconvert extended, critical, and localized sectors opens new opportunities for investigating mobility-edge physics, multifractality, non-ergodic dynamics, and programmable transport in coherent wave systems.

\noindent \textbf{Note added.} We note a related experimental work by Z. Hu \textit{et al.}~\cite{XJL}, which independently reported the observation of a quantum phase featuring coexisting extended, localized, and critical states.

\noindent \textbf{Acknowledgements.} 
J.F. acknowledges Prof. Xiong-Jun Liu for proposing the idea and initiating the project on the seven-phase landscape of Anderson localization, and for preliminary discussions with Dr. Xin-Chi Zhou. This work is supported by the Innovation Program for Quantum Science and Technology (2021ZD0300804), and the Key-Area Research and Development Program of Guangdong Province Grant No.2020B0303010001, Grant No.2019ZT08X324, No.2019CX01X042. Y. W. acknowledges support from National Key R\&D Program of China under Grant No. 2022YFA1405800.

\section{Appendix}

We present an illustration of the Floquet model and its effective Hamiltonian, together with the detailed procedures for identifying the seven distinct phases and constructing the phase diagram, as well as the experimental implementation details of the Floquet model.

\noindent \textbf{Illustration of the Floquet model and its experimental realization.}
The Floquet evolution over one driving period is defined by
$\mathcal{F}_{\rm OBC}=U_4U_3U_2U_1$,
whose implementation is illustrated in Fig.~\ref{fig4}. The operator $U_1$ performs a site-resolved rotation in the two-component spin space. The operators $U_2$ and $U_3$ each consist of a spin-dependent translation, a local spin rotation, and the inverse translation. Because the translation operator shifts only one spin component, these operations effectively convert internal spin rotations into directional nearest-neighbor hopping processes. The final step, $U_4$, applies staggered spin-dependent onsite phases, completing the Floquet driving cycle.

The experimental setup employs a recirculating fiber loop to emulate the Floquet lattice~\cite{PQW1,PQW2,PQW3,PQW4,PQW5,PQW6}. A 6-ns laser pulse (1560 nm) is injected into the loop via a 90:10 beam splitter (BS). The pulse first passes through a half-wave plate (HWP1) followed by an electro-optic modulator (EOM1). EOM1 applies a site-dependent phase shift via synchronized nanosecond electric pulses. The combination of HWP1 and the EOM1 realizes the rotation matrix $O(\theta_n) = e^{-i\theta_n\sigma_y}$ with $\theta_n = (-1)^n J/2$, implementing the onsite spin rotation $U_1$.

The translation-assisted hopping operations $U_2$ and $U_3$ are realized using a phase-stable double-loop interferometer. Upon entering via PBS1, the horizontal polarization ($|H\rangle$, corresponding to $|\uparrow\rangle$) component propagates through a 16-m single-mode fiber (long arm), while the vertical ($|V\rangle$, $|\downarrow\rangle$) component traverses a short arm. The two components recombine at PBS2, with the $|H\rangle$ path acquiring an 80-ns delay relative to $|V\rangle$. With the time-bin size set to $\tau_b = 80$ ns, this realizes the shift operator $S = \sum_n (|n+1,\uparrow\rangle\langle n,\uparrow| + |n,\downarrow\rangle\langle n,\downarrow|)$. After PBS2, the pulse passes through EOM2 and HWP3, which jointly implement a spin rotation $O(\theta_n') = e^{-i\theta_n'\sigma_y}$. For the first round trip ($U_2$), $\theta_n' = \theta_n^{(+)}$; for the second round trip ($U_3$), $\theta_n' = -\theta_n^{(-)}$. The two polarization components are then routed back through PBS2: now the $|H\rangle$ component takes the short arm and $|V\rangle$ the long arm, swapping the paths relative to the forward pass. This reverses the initial displacement, realizing $S^{-1}$ upon exiting PBS1. The net effect converts the internal spin rotation into directional nearest-neighbor hopping.  To realize this distinction, namely that $U_2$ applies $S$ followed by $S^{-1}$ whereas $U_3$ applies $S^{-1}$ followed by $S$, a HWP2 oriented at $45^\circ$ is inserted in the interferometer loop. 

The staggered onsite phase $U_4$ is implemented after the double-loop interferometer in the second round trip. The pulse passes through EOM3, placed between two quarter-wave plates (QWP1 and QWP2) and configured to apply a spin-dependent phase. The resulting operation is $U_4 = \sum_n |n\rangle\langle n| \otimes e^{i(-1)^{n+1}\phi\sigma_z}$ with $\phi = [10.4 - 1.3(\lambda_+ + \lambda_-)]J$.

A 4-km single-mode fiber delays the pulse by approximately $20~\mu$s, preventing temporal overlap between successive round trips. After each round trip, $10\%$ of the optical power is extracted by a beam splitter (BS) and directed to  single-photon detectors (SPDs). Detection events are time-stamped, enabling reconstruction of the site-resolved probability distribution $p(n,\uparrow(\downarrow),t)$. An erbium-doped fiber amplifier (EDFA) for round-trip losses. All site- and time-dependent modulations are programmed through synchronized electrical pulse sequences applied to EOM1,EOM2 and EOM3. 

\begin{figure}[t]
	\centering
	\includegraphics[width=0.8\columnwidth]{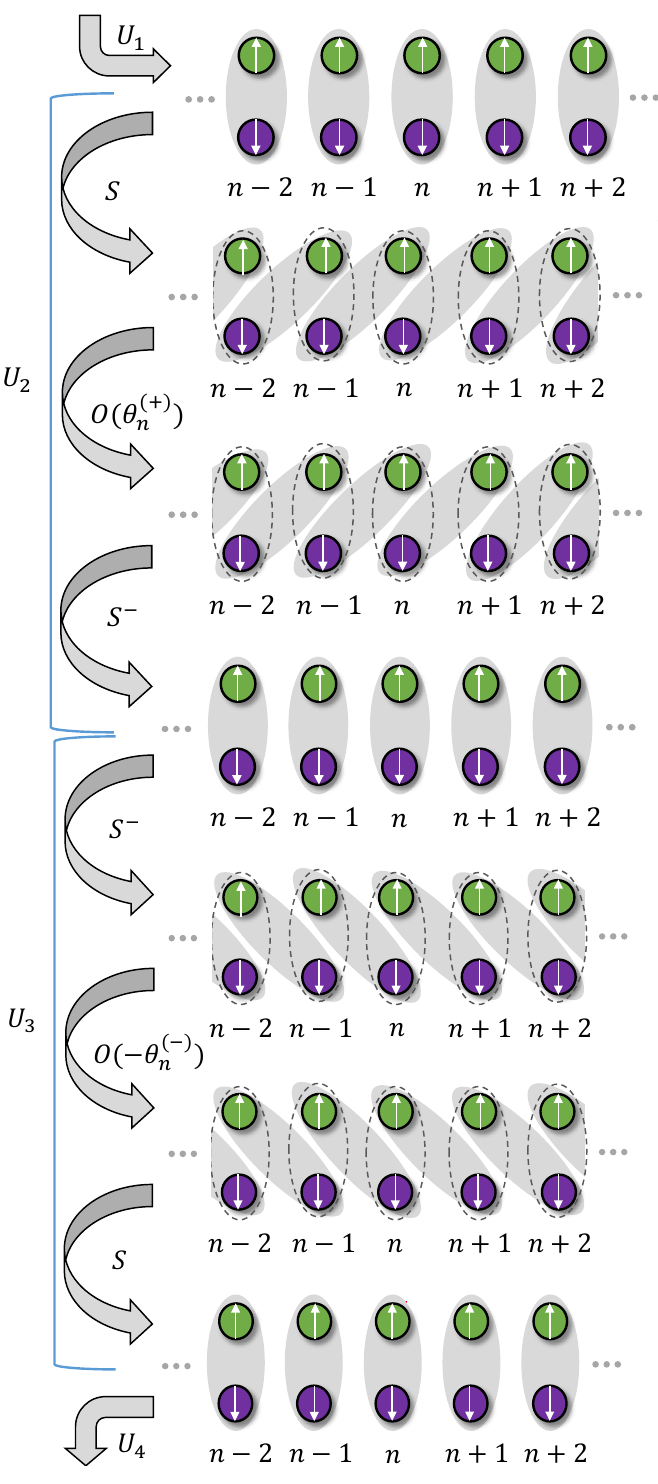}
	\caption{
    Schematic illustration of the four-step Floquet evolution $\mathcal{F}_{\rm OBC}=U_4 U_3 U_2 U_1$.
        }
	\label{fig4}
\end{figure}

\noindent \textbf{Effective Hamiltonian.} 
In the weak-driving limit, the Floquet Hamiltonian is approximated by the leading-order Baker-Campbell-Hausdorff (BCH) expansion,
\[
H_F = {\mathrm{i}}\ln(\mathcal{F}_{\rm OBC}) \approx H_1 + H_2 + H_3 + H_4,
\]
where \(H_j = {\mathrm{i}}\ln U_j\).

The first operator \(U_1 = \sum_{n=1}^N |n\rangle\langle n| \otimes O(\theta_n)\) is block-diagonal in the site basis, where the on-site rotation matrix is \(O(\theta_n) =\begin{bmatrix} \cos\theta_n & -\sin\theta_n \\ \sin\theta_n & \cos\theta_n \end{bmatrix}= e^{-{\mathrm{i}}\theta_n \sigma_y}\).
Its logarithm gives $H_1 = {\mathrm{i}}\ln U_1 = \sum_{n=1}^N \theta_n |n\rangle\langle n| \otimes \sigma_y$.
Using \(\sigma_y = -{\mathrm{i}}(|\uparrow\rangle\langle\downarrow| - |\downarrow\rangle\langle\uparrow|)\),
\[
H_1 = -{\mathrm{i}}\sum_{n=1}^N \theta_n \big(|n,\uparrow\rangle\langle n,\downarrow| - |n,\downarrow\rangle\langle n,\uparrow|\big),
\]
which represents an on-site spin rotation with site-dependent angle \(\theta_n\).

The operators \(U_2\) and \(U_3\) describe translation-assisted spin rotations, which convert spin flips into directional hopping.
We write \(U_2 = S^{-1} M S\) with
\(M = |1\rangle\langle 1| \otimes {\mathbb{I}} + \sum_{n=2}^N |n\rangle\langle n| \otimes O(\theta_{n-1}^{(+)})\).
Using \(\ln(S^{-1} M S) = S^{-1}(\ln M)S\), we have \(H_2 = {\mathrm{i}}\ln U_2 = S^{-1}({\mathrm{i}}\ln M)S\).
The matrix \(M\) is block-diagonal:
for \(n=1\), \(\langle 1|M|1\rangle = {\mathbb{I}}\); for \(n \ge 2\), \(\langle n|M|n\rangle = e^{-{\mathrm{i}}\theta_{n-1}^{(+)}\sigma_y}\).
Thus \(i\ln M = \sum_{n=2}^N \theta_{n-1}^{(+)} |n\rangle\langle n| \otimes \sigma_y
= \sum_{n=1}^{N-1} \theta_{n}^{(+)} |n+1\rangle\langle n+1| \otimes \sigma_y\).
The shift operator is
\(S = \sum_{n=1}^N \big(|n+1,\uparrow\rangle\langle n,\uparrow| + |n,\downarrow\rangle\langle n,\downarrow|\big)\)
(with \(|0,\uparrow\rangle \equiv 0\) under open boundaries).
Applying the similarity transform \(S^{-1}(\cdots)S\), yielding
\[
H_2 = -{\mathrm{i}}\sum_{n=1}^{N-1} \theta_{n}^{(+)} \big(|n,\uparrow\rangle\langle n+1,\downarrow| - |n+1,\downarrow\rangle\langle n,\uparrow|\big).
\]
A similar calculation for $U_3=S[\sum_{n=1}^{N-1}\ket{n}\bra{n}\otimes O(-\theta_{n}^{(-)})
+\ket{N}\bra{N}\otimes\mathbb{I}]S^{-1}$
gives
\[
H_3 = -{\mathrm{i}}\sum_{n=1}^{N-1} \theta_{n}^{(-)} \big(|n,\downarrow\rangle\langle n+1,\uparrow| - |n+1,\uparrow\rangle\langle n,\downarrow|\big).
\]

Finally, \(U_4 = \sum_{n=1}^N |n\rangle\langle n| \otimes e^{i(-1)^{n+1}\phi \sigma_z}\) imprints a staggered spin-dependent onsite phase.
Its logarithm yields
\[
H_4 = {\mathrm{i}}\ln U_4 = \sum_{n=1}^N (-1)^{n}\phi \big(|n,\uparrow\rangle\langle n,\uparrow| - |n,\downarrow\rangle\langle n,\downarrow|\big).
\]

Combining the four generators, we obtain the effective Hamiltonian 
$H_F \approx H_1 + H_2 + H_3 + H_4$.
Although the BCH expansion involves higher-order commutators
that are neglected in this approximation, we have verified that the phase diagram
obtained from ${\mathrm{i}}\ln\mathcal{F}_{\rm OBC}$ is consistent with that obtained from $H_F$
for the parameter ranges studied. Thus $H_F$ reliably captures the localization
physics of the Floquet system.\\ 
%After reordering the basis into two interlaced chains (see Supplementary Material), this Hamiltonian maps onto a ladder model with alternating hoppings, interchain coupling, and staggered potentials. \\

\begin{figure*}[t]
    \centering
    \includegraphics[width=1\textwidth]{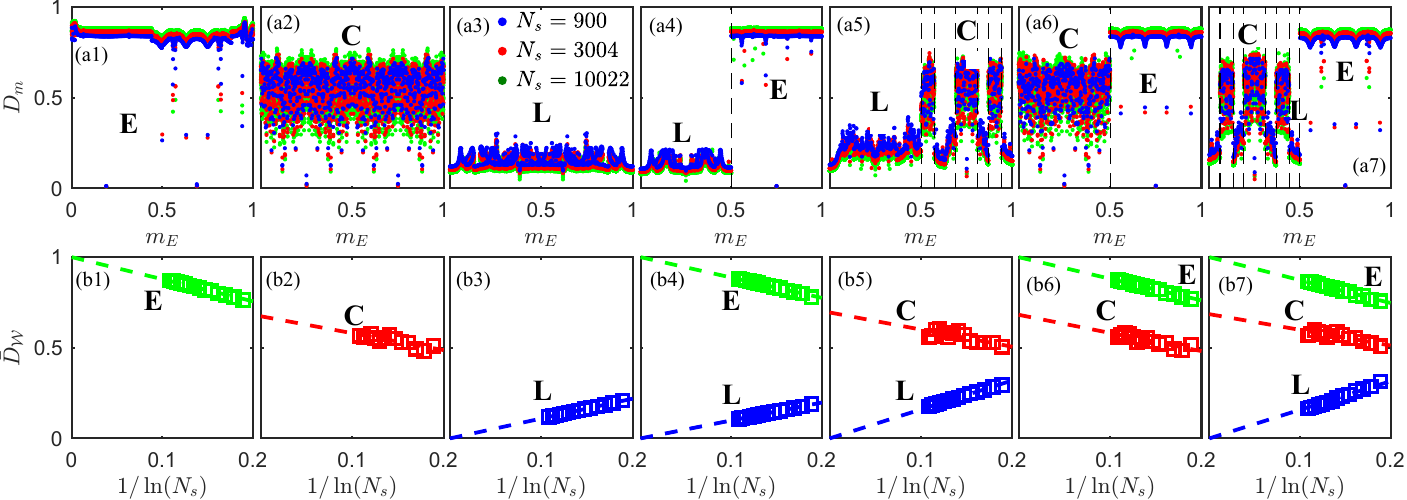}
    \caption{
    Finite-size scaling analysis of the fractal dimension for the seven
    representative phases. From left to right, the panels correspond to \((\lambda_+,\lambda_-)=(0,0.35), (1.1,1.1), (3.6,3.6), (3.6,0),(2.6,3.3), 
    (1.1,0.1), (2.3,0.2)\). Upper panels (a1)--(a7): Fractal
dimension Dm of each eigenstate versus the normalized quasienergy index $m_E=m/N_s$ for three system sizes $N_s= 1402, 3784$,
and $7962$. The dashed vertical lines delimit the quasienergy regions selected for further scaling analysis. The labels $E$, $C$, and
$L$ indicate the spectral regions identified as extended, critical, and localized, respectively. Lower panels (b1)--(b7): Scaling of the mean fractal dimension
    \(\overline{D}_{\mathcal{W}}\) as a function of \(1/\ln N_s\) for each selected
    region \(\mathcal{W}\) marked in the upper panels.
    Here \(\overline{D}_{\mathcal{W}}\) is the average of \(D_m\) over all eigenstates
    within \(\mathcal{W}\).
    Blue, red, and green curves and markers denote extended, critical, and
    localized components, respectively. 
    }
    \label{fig:scaling}
\end{figure*}

\noindent \textbf{Phase diagram via fractal dimension and finite-size scaling.}
The phase diagram is determined from the finite-size scaling of the fractal dimensions of the eigenstates obtained from ${\mathrm{i}}\ln(\mathcal{F}_{\rm OBC})$.
The \(m\)-th eigenstate is written as
$|\Psi_m\rangle = \sum_{n=1}^{N} \left( \psi_{m,n,\uparrow} |n,\uparrow\rangle + \psi_{m,n,\downarrow} |n,\downarrow\rangle \right)$,
with the total system size \(N_s = 2N\) accounting for the two spin components.
For each eigenstate, we evaluate the inverse participation ratio (IPR)~\cite{Anderson4}
$\mathrm{IPR}_m = \sum_{n=1}^{N} \left( |\psi_{m,n,\uparrow}|^4 + |\psi_{m,n,\downarrow}|^4 \right)$,
from which we obtain the finite-size fractal dimension
$D_m = - {\ln(\mathrm{IPR}_m)}/{\ln N_s}$.
Thermodynamically, \(D_m=1\), \(0\), and \(0<D_m<1\) mark extended, localized, and critical states, respectively.
For finite systems, the scaling of \(D_m\) with system size provides a robust diagnostic:
\(D_m\) increases with \(N_s\) for extended states, decreases for localized states, and remains nearly size-independent for critical states.
Figures~\ref{fig:scaling}(a1)--(a7) show \(D_m\) for all eigenstates at different system sizes across the seven phases.
In the pure extended phase [Fig.~\ref{fig:scaling}(a1)], \(D_m\) increases with system size; in the pure localized phase [Fig.~\ref{fig:scaling}(a2)], it decreases, and in the pure critical phase [Fig.~\ref{fig:scaling}(a3)], it exhibits size-independent fluctuations. For the extended-localized mixed phase [Fig.~\ref{fig:scaling}(a4)], some eigenstates show increasing \(D_m\) and others decreasing \(D_m\). For the critical-localized mixed phase [Fig.~\ref{fig:scaling}(a5)], part of the spectrum shows decreasing \(D_m\) while the remainder exhibits size-independent fluctuations.
For the extended-critical mixed phase [Fig.~\ref{fig:scaling}(a6)], part of the spectrum shows increasing \(D_m\) while the remainder exhibits size-independent fluctuations.
For the phase where all three types coexist [Fig.~\ref{fig:scaling}(a7)], regions of increasing, decreasing, and size-independent \(D_m\) are simultaneously present across the spectrum.

The regions marked by \(E\), \(C\), and \(L\) in the upper panels of Fig.~\ref{fig:scaling} are selected for finite-size scaling analysis, 
the results of which are shown in 
Figs.~\ref{fig:scaling}(b1)--(b7).
For each selected quasienergy window \(\mathcal{W}\) (delimited by the dashed 
vertical lines in the upper panels), we compute the mean fractal dimension
\[
\overline{D}_{\mathcal{W}}(N_s) = \frac{1}{N_{\mathcal{W}}} \sum_{m\in\mathcal{W}} D_m(N_s),
\]
where \(N_{\mathcal{W}}\) is the number of eigenstates contained in \(\mathcal{W}\).
That is, \(\overline{D}_{\mathcal{W}}\) is the average of the fractal dimensions 
of all eigenstates whose quasienergies fall into \(\mathcal{W}\).
Figures~\ref{fig:scaling}(b1)--(b7) plot \(\overline{D}_{\mathcal{W}}\) against 
\(1/\ln N_s\) for each selected window.
For pure phases, the entire spectrum converges to a single limiting value:
\(\overline{D}_{\mathcal{W}} \to 1\) in the extended phase, 
\(\overline{D}_{\mathcal{W}} \to 0\) in the localized phase, and 
\(\overline{D}_{\mathcal{W}} \to D_c \in (0,1)\) in the critical phase.
For mixed phases, different spectral windows exhibit distinct asymptotic 
behaviors, reflecting the coexistence of multiple transport sectors. \\

\end{document}